\documentclass{ws-procs975x65}

\begin{document}

\title{MICROSCOPIC STUDIES OF SOLID $^4$HE
WITH PATH INTEGRAL PROJECTOR MONTE CARLO}

\author{M. ROSSI, R. ROTA, E. VITALI, D.E. GALLI$^*$ and L. REATTO}

\address{Dipartimento di Fisica, Universit\`a degli Studi di Milano,\\
via Celoria 16, 20133 Milano, Italy\\
$^*$E-mail: Davide.Galli@unimi.it}

\begin{abstract}
We have investigated the ground state properties of solid $^4$He with the Shadow Path Integral 
Ground State method.
This exact $T=0$ K projector method allows to describes quantum solids without introducing any 
{\it a priori} equilibrium position.
We have found that the efficiency in computing off-diagonal properties in the solid phase 
sensibly improves when the direct sampling of permutations, in principle not required, is 
introduced.
We have computed the exact one-body density matrix (obdm) in large commensurate $^4$He crystal 
finding a decreasing condensate fraction with increasing imaginary time of projection, making our 
result not conclusive on the presence of Bose-Einstein condensation in bulk solid $^4$He.
We can only give an upper bound of $2.5\times10^{-8}$ on the condensate fraction.
We have exploited the SPIGS method to study also $^4$He crystal containing grain boundaries by 
computing the related surface energy and the obdm along these defects.
We have found that also highly symmetrical grain boundaries have a finite condensate fraction.
We have also derived a route for the estimation of the true equilibrium concentration of vacancies 
$x_v$ in bulk $T=0$ k solid $^4$He, which is shown to be finite, $x_v=(1.4\pm0.1)\times10^{-3}$ at
the melting density, when computed with the variational shadow wave function technique.
\end{abstract}


\bodymatter

\section{Introduction}

The low-temperature physics of $^4$He samples has been a continuous test ground of many-body 
theories; though displaying all the features of a strongly interacting system, a collection of 
$^4$He atoms has a very simple effective Hamiltonian: a model of spinless and structureless bosons 
interacting through a two-body interaction potential has been proved to provide a very accurate 
description of the properties of liquid and solid Helium, so that, when dealing with $^4$He, one 
does not have to face the added complexities related to Fermi statistics or to nuclear Physics.
The phenomenon of superfluidity in liquid Helium, displaying in a macroscopic domain the quantum
laws of nature related to the boson indistinguishability of the atoms, has been the object of 
several decades of theoretical efforts; the development of very accurate computer simulations 
methods has provided the possibility to put light into the physical mechanisms which govern such 
striking experimental observations.

It is known that, under pressure, at very low temperature, a sample of Helium atoms solidifies.
For many years it has been argued that quantum coherence phenomena, such as Bose Einstein 
Condensation (BEC) and superfluidity, could take place even in the crystalline phase if the 
compromise between particles localization induced by the crystalline order and particles 
delocalization due to the zero-point motion of the atoms allowed the boson nature of the particles 
to play an important role giving rise to a supersolid phase.
The supersolid state of matter, formerly appeared only in the speculations of some eminent 
theoreticians\cite{andr,ches,leggett}, has been elusive for decades\cite{meis} up to its first 
probable evidence in torsional oscillator experiments with solid $^4$He\cite{chan}.
Now, after three years of renewed interest and intense experimental and theoretical studies, 
physicists still find themselves in front of a puzzle of signs difficult to recompose\cite{prot}.
Non classical rotational inertia (NCRI) effects in torsional oscillator experiments with solid 
$^4$He (which have been observed also in confined samples\cite{cha2}) have been observed in a number
of laboratory\cite{repp,shir,kubo,koji} and recently also the robustness of these effects has been 
tested by working with single crystals\cite{cha3}.
The whole experimental scenario reached after three years of research activity is not capable to 
clearly discern whether such effects arise from the true thermal equilibrium state of the system or
whether they are due to non equilibrium properties.
From one side, if one omits the amazing direct observation of superfluid effects in presence of 
grain boundaries\cite{Balibar} and the specific heat peak detected near 80 $m$K in solid 
$^4$He\cite{cha3}, NCRI effects in torsional oscillator are still the only manifestation of the 
elusive supersolid phase; from the other side it is clear that such effects are strongly affected by
disorder but what is not clear is to what extent the disorder plays a role, because there are 
conflicting conclusions on the effects of annealing processes and now there is also evidence of a 
variable amount of NCRI in different samples even in single crystals after annealing\cite{cha3}.

Very accurate ``ab initio'' microscopic simulation methods, whose importance in exploring the liquid
phase has been outstanding, could give a deep insight into solid Helium physics. 
We have used the Shadow Path Integral Ground State (SPIGS) method\cite{spigs} to investigate the 
zero-temperature properties of solid $^4$He.

In the first section we introduce the SPIGS technique; the evaluation of the one body density matrix
in both a commensurate and incommensurate sample of solid Helium is reported in the second section.
Then we deal with defects, vacancies and grain-boundaries: we describe a method to evaluate the
concentration of zero-point vacancies, which are known to help the delocalization of the atoms thus
favoring BEC in the system, and to study their dynamics, and, in the following section, we report 
our diagonal and off-diagonal results in presence of grain-boundaries.

\section{Path Integral Projector Monte Carlo methods}

The Path Integral Ground State Method (PIGS) is a $T=0$ K projector Quantum Monte Carlo method which
allows the calculation of ``exact'' ground state averages in a quantum system using a wave function
$\Psi_\tau$ that asymptotically approaches the unknown ground state $\Psi_0$\cite{pigs}.
$\Psi_\tau$ is obtained via iterative applications of the operator $e^{-\delta \hat{H}}$ (this 
action will be called {\it projection} in the following) on an initial trial state, $\Psi_T$, 
characterized by a non-zero overlap with the true ground state $\Psi_0$:
$\Psi_\tau=(e^{-\delta\hat{H}})^P\Psi_T=e^{-\tau\hat{H}}\Psi_T$,
where $\hat H$ is the Hamiltonian operator of the quantum system and $\delta=\tau/P$.
The projection procedure exponentially removes (as a function of $\tau$) from $\Psi_\tau$ any 
overlap of $\Psi_T$ with the excited states.
In the position representation the corresponding wave function 
$\Psi_\tau(R)=\langle R|\Psi_\tau\rangle$ is expressed as a succession of $P$ convolutions with the 
imaginary time propagator $G(R,R',\delta)=\langle R|e^{-\delta\hat{H}}|R'\rangle$ to give the Path 
Integral approximation of the ground state wave function:
\begin{equation}\label{psi_spigs}
 \Psi_\tau(R)=\int\prod_{j=1}^{P} [dR_j \, G(R_{j+1},R_j,\delta)] \, \Psi_T(R_1)
\end{equation}
where $R_{P+1}\equiv R$, $\Psi_T(R)=\langle R | \Psi_T \rangle$ and generally 
$R=\{ \vec{r}_1,\vec{r}_2,...,\vec{r}_N \}$ represents a set of coordinates for the $N$ particles 
in the quantum system.
If $\delta$ is chosen sufficiently small, accurate approximations for $G(R,R',\delta)$ are known
even for a strongly interacting system like a collection of $^4$He atoms in the solid 
phase\cite{ceperley}.
When only one projection step is considered and the imaginary time propagator is approximated via a
variationally optimized ``primitive'' density matrix\cite{ceperley}, the well known variational shadow 
wave function (SWF) technique is recovered\cite{swf,moro}.
With SWF, which is translationally invariant and contains implicitly correlations at all the orders
in the number of particles, the solid phase emerges as a result of a spontaneously broken 
translational symmetry\cite{swf}.
This is done in a so efficient way that it is possible to describe the solid and the liquid phase 
with the same functional form since there is no need of {\it a priori} known equilibrium positions 
for the crystalline phase.
Moreover SWF presently represents the best variational description of liquid and solid 
Helium\cite{moro} in the sense that it gives the lowest energy in both the phases.

The Shadow Path Integral Ground State method (SPIGS) corresponds to a PIGS method in which the 
projection procedure is applied to a SWF\cite{spigs}.
Due to the fact that the first imaginary time projection step is done with the SWF technique, SPIGS
method allows to describe the solid phase without explicitly breaking the translational symmetry.
To our knowledge SPIGS is the only projector quantum Monte Carlo method which provides the 
possibility to study disorder phenomena and the effects of the indistinguishability of the particles
in a quantum solid at $T=0$ K\cite{spigs}.

The Monte Carlo methods for the calculation of multi-dimensional integrals can be used with PIGS and
SPIGS because it is easy to show that expectation values computed with the wave function in 
\eqref{psi_spigs} are formally equivalent to canonical averages of a classical system of special 
interacting linear open polymers\cite{pigs,spigs}.
This is very similar to what is done in a PIMC simulation, where finite temperature (and not $T=0$ 
K) averages are computed and these are found to be equivalent to canonical averages for a classical
system of ring polymers\cite{ceperley}.
The calculation of expectation values with \eqref{psi_spigs} will differ from the true ground state
averages by a quantity that can be reduced below the statistical error of the MC calculation by an
appropriate choice of $\delta$ and the number of projections $P$, or equivalently of $\tau=P\delta$.
This is the reason why we can talk about ``exact'' projection techniques in relation with PIGS and 
SPIGS, and why with these methods the convergence with $\tau$ of the computed expectation values 
should be always checked.

\section{One-Body density matrix in solid $^4$He}

The appropriate Monte Carlo calculations of statistical averages for a Boson system at {\it finite 
temperature} with the PIMC method require the sampling of permutations between particles in order
to fulfill the Bose symmetry\cite{ceperley}.
The fact that without such procedure one cannot recover the correct results can be easily understood
by thinking about the classical isomorphism: the sampling of permutations between particles leads to 
configurations in which more ring polymers join together into a single one; such configurations,
which are topologically unreachable without permutations, are of primary importance in computing 
off-diagonal and superfluid properties because such configurations are those which contribute to 
give a non-zero superfluid fraction when computed with the winding number estimator\cite{ceperley}.

The situation is different with the Path Integral projector methods. 
The projection procedure applied to a wave function which is Bose symmetric, protects the Bose 
symmetry in $\Psi_\tau$.
This can be easily perceived by thinking about the different classical isomorphism which 
characterizes these $T=0$ K methods: the corresponding classical system is made of linear open 
polymers; the sampling of permutations between particles corresponds to enable linear polymers to 
mutually exchange one of their two tails, leaving the system in a configuration which is not 
topologically different form the starting one and that can be reached by a combination of some 
(perhaps many) simple moves that involve a single polymer one at a time.
While permutations moves are not necessary in SPIGS, it is possible however that such moves in some
situation could dramatically improve the efficiency of the sampling procedure by assuring the 
effective ergodicity of the algorithm.
This is why recently we have introduced also their sampling in SPIGS; the sampling scheme that we 
have implemented in our SPIGS algorithm is the one reported in Ref.~\refcite{boninsegni}.
We have verified that there are no substantial improvements in the calculations of diagonal 
properties both in the liquid and in the solid phase; however permutations turn out to be very 
important for a really efficient estimation of non diagonal properties in the solid phase 
such as the one body density matrix (OBDM), as we shall discuss below.

The OBDM $\rho_1 (\vec{r} - \vec{r}\,')$, at zero temperature, is defined as follows:
\begin{equation}
 \label{rodlro}
 \rho_1(\vec{r}-\vec{r}\,')=N\int\!{d\vec{r}_2,\dots,d\vec{r}_N}\,
                                \Psi_0(\vec{r},\vec{r}_2,\dots,\vec{r}_N)
                                \Psi_0(\vec{r}\,',\vec{r}_2,\dots,\vec{r}_N)
\end{equation}
where $\Psi_0$ is the ground state wave function.
From the behavior of $\rho_1(\vec{r}-\vec{r}\,')$, for $|\vec{r}-\vec{r}\,'|\rightarrow\infty$, 
which is directly related to the fraction of atoms occupying the zero-momentum one particle state, 
one can infer whether the system exhibit BEC.
The known property that, at $T=0$ K, BEC implies non classical rotational inertia, makes extremely
important the evaluation of the OBDM in solid Helium.

The reality and positivity of the ground state wave function, but also of its approximation 
$\Psi_{\tau}$, allows one to interpret the integrand in \eqref{rodlro} as a canonical weight of a 
classical system of polymers, one of which has been cut into two halves (we will call them {\it half
polymers}), one connected to the point $\vec{r}$ and the other connected to the point $\vec{r}\,'$.
Therefore $\rho_1$ turns out to be the histogram of the relative distance of this two points which 
can be sampled, together with the other coordinates, using a standard Metropolis algorithm.

When computing the OBDM in the solid phase, the sampling of configurations in which the two half
polymers are far away faces ergodicity problems due to the potential barriers arising from hard core
interactions; in fact, the polymers are shown to have an high probability to be rolled up around the
equilibrium positions of the lattice.
In this framework an accepted permutation, which involves one of the two half polymers with the 
other polymers, allows it to go beyond, at least partially, an adjacent polymer ($^4$He atom) and 
then to overwhelm the potential barriers which slow down the ergodicity of the sampling.
During a calculation of the OBDM in solid $^4$He at the melting density, this extended SPIGS
algorithm is able to sample a permutation involving one half polymer about every 100 MC steps
without losing computational efficiency.
We have found that a simulation run of about 10$^6$ MC step is long enough to observe permutation
cycles which involve one of the two half polymers plus a number of other polymers up to 6-7.
Permutations between two and three polymers represent the 80\% and 16\% of the sampled cycles 
respectively; adding one more polymer to the cycle is found to reduce by an order of magnitude the
relative frequency of accepted permutation cycle.

\begin{figure}[t]
 \label{obdm}
 \centerline{\psfig{file=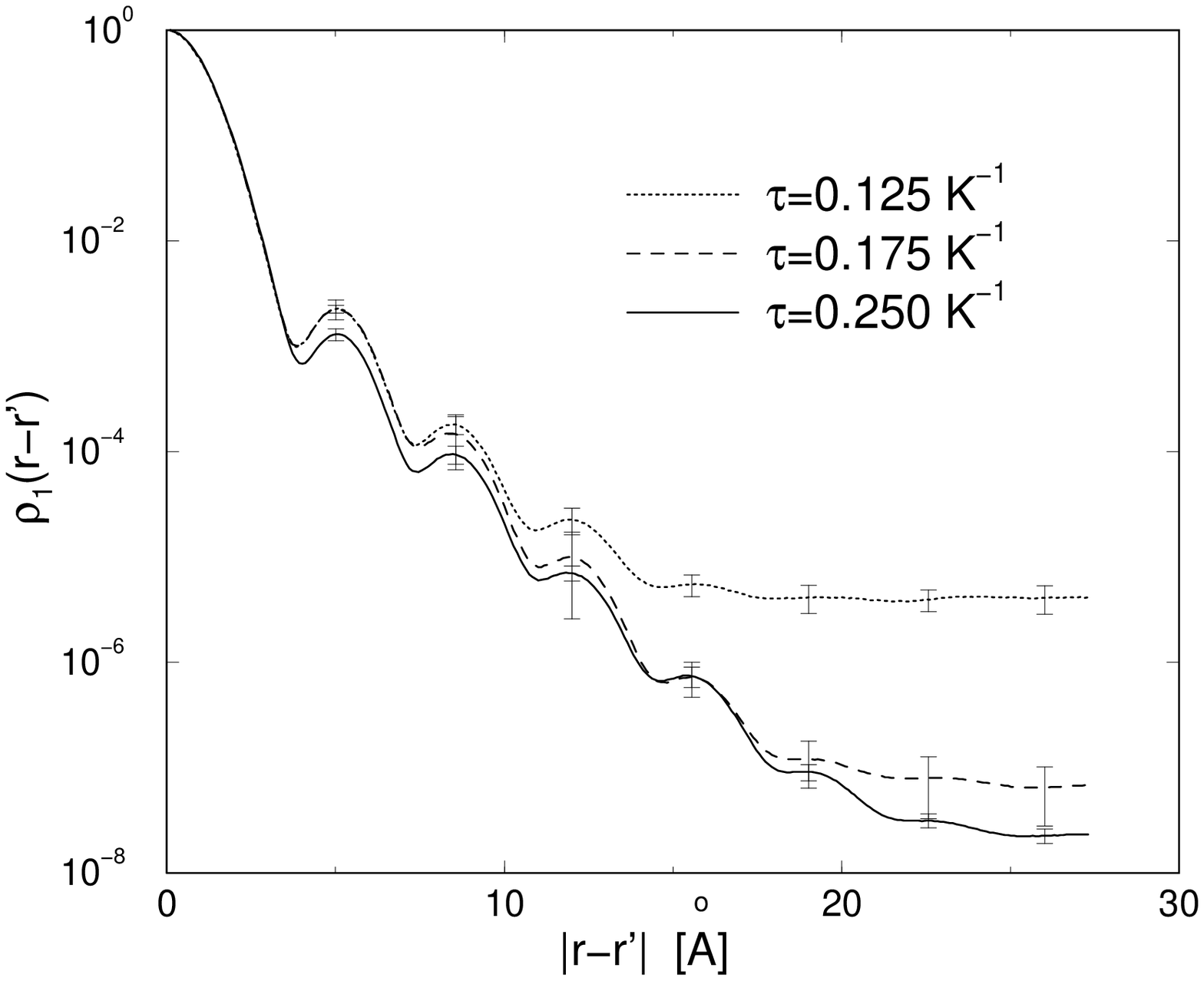,width=5.8cm}{(a)}
             \psfig{file=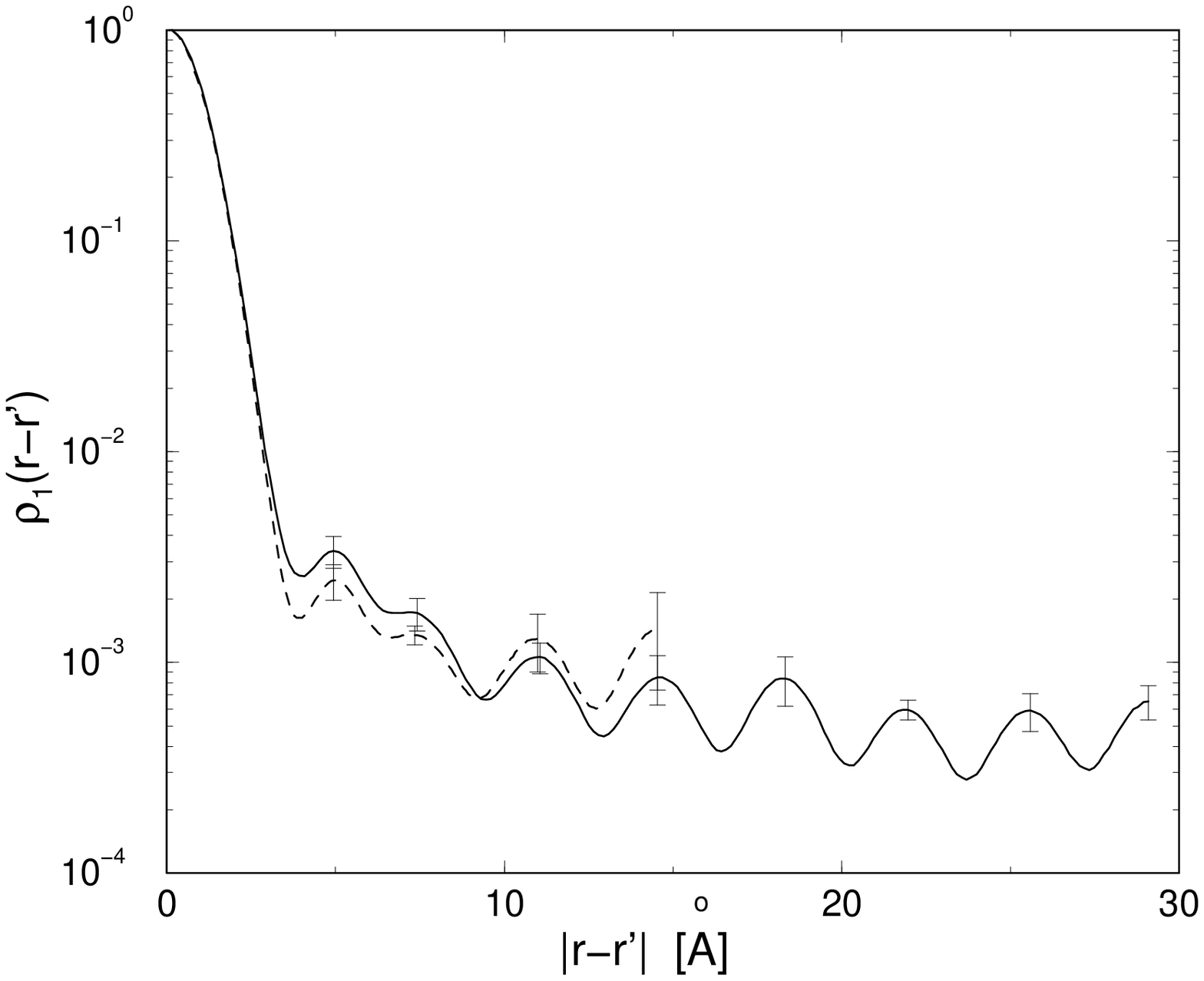, width=5.8cm}{(b)}}
 \caption{One body density matrix computed along the nears neighbor direction in the basal plane of
          a $^4$He hcp crystal at $\rho=0.0293$ \AA$^{-3}$: (a) commensurate solid (b) solids with 
          fixed concentration of vacancies $x_v=1/287$.}
\end{figure}
In Fig.~\ref{obdm} we show the OBDM computed in a commensurate hcp crystal and in a hcp crystal at 
fixed vacancy concentration $x_v=1/287$ at the melting density $\rho=0.0293$ \AA$^{-3}$.
The variational estimate of the condensate fraction $n_0$ performed with the SWF in the commensurate
crystal gave $n_0=(5.0\pm1.7)\times10^{-6}$\cite{gal2}.
From evaluations with the ``exact'' $T=0$ K SPIGS method, we have seen that this $n_0$ is remarkably
lowered by increasing the number of projection steps toward the true ground state; this presently 
allow us to provide only an upper bound of  about $2.5\times10^{-8}$ for the condensate fraction in
the commensurate crystal.
We can argue that, in relation with the computation of the OBDM in the commensurate crystal, some 
correlations are still missing in the variational SWF, resulting in a $n_0$ value far away from the 
true one, which is recovered only after many projection steps. 
This limitation of the SWF was not observed when a finite concentration of vacancies was present in 
the system\cite{galn}; in this case, in fact, the condensate fraction was shown to converge quickly 
to the true value after few projections.
With the aid of the new extended SPIGS we were able to efficiently study the behavior of $\rho_1$ in
large hcp crystals with a finite concentration of vacancies.
As shown in Fig.~\ref{obdm}(b), the large distance plateau in the OBDM tail clearly indicates the 
presence of ODLRO in the system with vacancies; and, by keeping $x_v$ constant while enlarging the 
system from 287 to 574 particles, we cannot detect any remarkable size dependence of $n_0$.

\section{Equilibrium concentration of vacancies}

One of the fundamental questions which has not yet an answer regards the true nature, commensurate 
or incommensurate, of the ground state of solid $^4$He\cite{galn}. 
By commensurate we mean that the number of atoms $N$ is equal to the number of lattice sites $M$, by
incommensurate that $N\neq M$.
Present experiments\cite{simm} do not give any direct evidence for zero point vacancies, i.e. 
vacancies in the ground state of solid $^4$He, but they are able only to put an upper bound of about
0.4\%\cite{simm2} on their equilibrium concentration at low temperatures.
Neither microscopic computations present in literature do allow to directly answer the 
question\cite{galn}, even if arguments against zero point vacancies other than metastable defects 
have recently appeared\cite{bonin2}.
In fact, in Monte Carlo simulations of crystals, periodic boundary conditions (pbc) and lattice 
structure impose a constraint that makes impossible, in practice, for the system to develop an 
equilibrium concentration of vacancies\cite{andersen}.
This holds not only for canonical and micro-canonical computations, but also for grand canonical 
simulations\cite{andersen}.
Furthermore the equilibrium concentration could be so small to easily have escaped detection in 
simulations of the ground state of solid $^4$He\cite{anderson}.
The equilibrium concentration of zero point vacancies $x_v$ can then be obtained only indirectly, by
a statistical thermodynamical analysis of an extended system, exactly as for classical 
solids\cite{frenkel}.

Following the Chester's argument\cite{ches}, we can infer that translationally invariant wave 
functions, as the Jastrow wave function (JWF) of the original Chester derivation, describe a solid 
with a finite $x_v$\cite{galn}.
This argument can be extended also to the SWF and to the SPIGS one.
As we have already pointed out, there exists a direct isomorphism between the quantum system at $T=0$
K and a suitable classical one at a finite $T$.
It is known that classical solids have a finite equilibrium concentration of vacancies even if a 
single vacancy has a finite cost of local free energy because of the gain in configurational 
entropy\cite{kittel}.
Then, given a wave function which describes the ground state of solid $^4$He, the problem of 
computing the equilibrium concentration of vacancies, if any, is formally equivalent to the estimate
of $x_v$ in the corresponding equivalent classical solid (ecs).
The first calculation of $x_v$ for solid $^4$He based on this quantum-classical isomorphism was 
done few years ago\cite{still} for a JWF with a McMillan\cite{mcmi} pseudopotential.
However it is known that a JWF is not a good variational wave function for solid $^4$He\cite{moro}. 
We have now extended this route for computing $x_v$ to the SWF, which gives the best variational 
description of $^4$He\cite{moro}.

\begin{figure}[t]
 \label{gvv}
 \centerline{\psfig{file=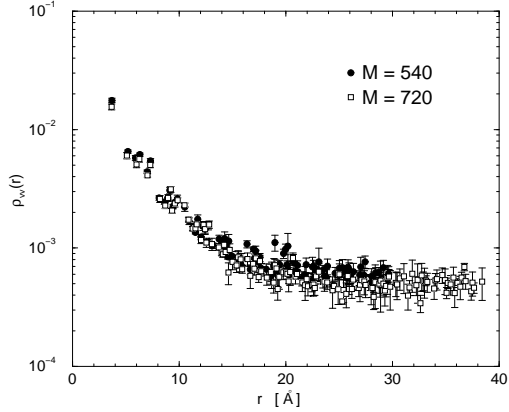,height=5.5cm}}
 \caption{Intervacancy correlation function $\rho_{\rm vv}(r)$ computed with the SPIGS method in 
          two hcp crystals at the melting density containing 3 vacancies.
          The value of $\rho_{\rm vv}(r)$ represent the probability of observing two vacancies
          separated by a distance $r$ normalized to the multiplicity of the distance $r$.
          Each $\rho_{\rm vv}(r)$ is obtained considering at least $4\times10^6$ configurations.}
\end{figure}
A preliminary aspect to be investigated is the interaction among vacancies.
Recent finite temperature Monte Carlo simulations\cite{bonin2} suggest the existence of some kind 
of attractive interaction, which makes already three vacancies form a tight bound state\cite{bonin2}
and the vacancy gas instable.
With the SPIGS method, we have computed the energy per particle $E$ in systems of increasing size 
which house an increasing number of vacancies ($M=180$, 360 and 540 with 1, 2 and 3 vacancies 
respectively) in such a way that the concentration of vacancies is preserved in order to rule out 
non-linear dependence effects on the concentration of vacancies.
In the $M=180$ case, the pbc constraint the vacancies to be separated by a distance equal to the box
side; then, if the attraction among vacancies play a relevant role there should be a gain in $E$ for
the large systems where vacancies can be close together: this gain would correspond to the binding 
energy.
Without tail corrections, the obtained energies are $E=-4.809\pm0.003$ for $M=180$, 
$E=-4.800\pm0.002$ for $M=360$ and $E=-4.804\pm0.002$ for $M=540$ with 1, 2 and 3 vacancies 
respectively.
Due to the chosen geometry for the simulation boxes, the tail corrections turn out to be the same 
in all the systems.
Since the results for $E$ are compatible within the statistical error, we argue that the interaction
among vacancies, if any, is very weak with a binding energy lower than the statistical error.
In order to give a deeper insight into such an interaction, we have studied also the intervacancy 
correlation function by monitoring their relative distance in the Monte Carlo sampling of 
incommensurate crystals constructing a radial correlation function $\rho_{\rm vv}(r)$.
From $\rho_{\rm vv}(r)$, plotted in Fig.~\ref{gvv} we see that the vacancies are always able to 
explore the whole available distance range. 
Moreover we observe a depletion in the short distance region when enlarging the system size. 
We may so conclude that, at least in the considered spatial dimensions, no sign of tight bound state
is detected. 
The observed large value of $\rho_{\rm vv}(r)$ at nearest neighbor distance is an indication of some
short range attraction, but this is not large enough to give a bound state at least for up to 3
vacancies.
We conclude that the vacancy gas should be stable in bulk solid $^4$He and, for very low 
concentrations as the expected equilibrium one, the approximation of vacancies as non interacting 
defects seems to be not so unreasonable.

A full grand canonical analysis of an extended classical solid, under the assumption of non 
interacting defects, leads to an equilibrium concentration of vacancies\cite{frenkel} 
$x_v= e^{-\beta(\mu-f_1)}$, where $\mu$ is the chemical potential of the perfect crystal, $-f_1$ is
the variation in the free energy associated with the formation of one vacancy and $\beta=1/k_BT$.
Exploiting thermodynamic relations, the exponent in $x_v$ can be written as 
$-\beta(\mu-f_1) = (M-1)\beta(f_0-f_d)-\beta P/\rho$.
Then, in order to compute $x_v$, we need the free energy per particle $\beta f_0$, and the pressure
$\beta P/\rho$ of the perfect ecs, and the free energy per particle $\beta f_d$, of the defected
(i.e. with one vacancy) ecs.
Let us consider the ecs of the SWF: it is a solid of triatomic molecules\cite{swf,moro}.
The pressure is quite straightforwardly obtained by the virial method\cite{fbook} or from volume 
perturbations\cite{volum}.
The computation of the free energy (both $\beta f_0$ and $\beta f_d$) is less immediate: it is 
related to the volume in the configurational space for the ecs (or to the normalization constant for
the quantum crystal which is never explicitly computed in a Monte Carlo simulation).
In classical systems, the thermodynamic integration method provides a way for reconstructing free 
energy differences by integration over a reversible path in the phase space\cite{fbook}.
For solids this is done by means of the Frenkel-Ladd method\cite{ladd,polson} (FLM), whose basic 
idea is to reversibly transform the solid under consideration into an Einstein crystal.
The implementation of FLM is not trivial, we will give more details on our computation method 
elsewhere\cite{nuovo}.

We have checked our numerical code for the estimate of the free energy in classical systems with the
FLM by reproducing the results in Ref.~\refcite{polson} for a system of soft spheres.
A check on the pressure is provided by the convergence to the same value of the two employed 
methods.
With the SWF, the equilibrium concentration of vacancies at the melting density 
$\rho=0.0293$\AA$^{-3}$ turns out to be $x_v=(1.4\pm0.1)\times10^{-3}$.
Since vacancies have been proved to be extremely efficient in inducing BEC in solid 
$^4$He\cite{galn}, this result would suggest the presence of a finite condensate fraction in bulk 
solid $^4$He, which results then in a supersolid.
Moreover, using $T_{\rm BEC}$ for an ideal Bose gas with mass equal to the vacancy effective mass 
$m_v=0.35m_{\rm He}$\cite{gal3} as an estimate of the supersolid transition temperature, we obtain 
$T\simeq11.3x_v^{2/3}=141 $ mK, which is about 2.3 times larger than the experimental transition 
temperature $T=60$ mK\cite{cha3}.
A possible origin of this small discrepancy is the fact that $x_v$ is estimated via a variational 
technique.
Fortunately this route for the estimate of the equilibrium concentration of vacancies $x_v$ can be 
applied also to the ``exact'' wave function \eqref{psi_spigs} resulting from the SPIGS method.
The calculation of $x_v$ as a function of $\tau$ is under way, we expect that the variational
result will be progressively modified under successive imaginary time projections until it reaches 
its true equilibrium value.

\section{Grain Boundaries}

The study of grain boundaries in solid helium have recently gained a particular interest after the 
direct experimental observation of superfluidity effects in crystals containing grain 
boundaries\cite{Balibar}.
Here we report a preliminary study on such defects in solid $^4$He by means of the SPIGS method; 
this is the first application of an ``exact'' $T=0$ K technique in the study of such systems.
We focused on the analysis of crystals containing highly symmetrical grain boundaries: in the system
that we will call SGB (from stable grain boundaries), the misorientation between the two 
crystallites is created by a relative rotation of the two lattices around a direction perpendicular
to the hcp basal planes (z-axis) of $\vartheta=2\arctan(\sqrt{3}/15)\simeq 13^o$; the number of 
particles ($N=456$) and the dimensions of the simulation box ($L_{x}=2a\sqrt{57}$, 
$L_{y}=a\sqrt{19}$, $L_{z}=2a\sqrt{6}$, being $a$ the \emph{lattice constant}) are chosen in order 
to match the periodicity of the hcp crystal. 
The application of pbc in each direction implies the presence of two planar grain boundaries which 
are perpendicular to the $x$ axis.
We have studied also other grain boundaries obtained with different relative rotations around the 
$z$-axis ($\vartheta\simeq32^o$ and $\vartheta\simeq38^o$) but also around other axis (the $y$ axis 
with $\vartheta\simeq22^o$).
The imaginary time parameters used in these simulation was $\delta=0.025$ K$^{-1}$ and $\tau=0.125$ 
K$^{-1}$ with the pair-product approximation for the imaginary time propagator; the test on the 
convergence with $\tau$ is under way.

Only the SGB system, among those we studied, showed stable grain boundaries as discussed in the 
following.
In our zero temperature simulations, we have observed that the grain boundaries are generally 
characterized by a high mobility: the positions of the two grain boundaries are found to move during
the simulations even of many {\AA}ngstroms in few thousand of Monte Carlo steps.
In the SGB system we have observed the two grain boundaries to move always in phase keeping 
essentially fixed their relative distance that was about 27 {\AA} at $\rho=0.0313$ \AA$^{-3}$.
For this reason, in the SGB system, the two crystallites with different orientations have a similar 
number of $^4$He atoms, even if the portion of the simulation cell that they occupy changes during 
the simulation run.
Among the other grain boundaries, those obtained by rotating the lattices around the $z$ axis 
manifest the same high mobility but, as opposed to the SGB system, the two interfaces show 
attractive or no correlations.
Due to this motion, we found that in few thousand of MC steps, the two grains collapse, leaving in 
the simulation box a single commensurate crystal.
Instead, the system obtained by rotating around the $y$ axis showed a rapid (few tens of thousands 
of MC steps) rearrangement of the atoms toward a highly stressed crystal without grain boundaries.
\begin{table}[t]
 \tbl{Surface energy of a symmetrical tilt grain boundary obtained with a rotation of 
      $\vartheta \simeq 13^o$ around the $c$-axis in an hcp crystal.}
 {\begin{tabular}{@{}ccccc@{}}
   \toprule
    &  & \multicolumn{3}{c}{Surface energy for systems}\\
    Density & Surface energy & \multicolumn{3}{c}{in presence of vacancies (K \AA$^{-2}$)}\\
    (\AA$^{-3}$)& (K \AA$^{-2}$)& 1 vac. & 2 vac. & 4 vac. \\
   \colrule
    0.0303 & $0.269 \pm 0.008$ & $0.244 \pm 0.006$ & GB unstable & - \\
    0.0313 & $0.305 \pm 0.004$ & $0.283 \pm 0.004$ & $0.250 \pm 0.005$ & GB unstable \\
    0.0333 & $0.406 \pm 0.008$ & $0.366 \pm 0.008$ & $0.325 \pm 0.008$ & - \\
    0.0353 & $0.546 \pm 0.005$ & $0.482 \pm 0.006$ & $0.418 \pm 0.005$ & $ 0.261 \pm 0.005$ \\
   \botrule
  \end{tabular}}
 \label{tabella}
\end{table}

We focused then our simulations on the SGB system.
In order to investigate on correlated motion of the two grain boundaries in the SGB system, we have
simulated it in a larger box in which the interfaces were placed at different distances in pbc:
we found that the two grain boundaries moved in order to maximize their distance in pbc, confirming
the repulsive intergrain interaction.
The observation at $T=0$ K of the high mobility and the correlated motion of the grain boundaries is
a surprising effect which is probably connected to a sort of recrystallization waves in systems with
such kind of defects.
For the SGB system we were able to compute the grain boundary surface energy $\cal E$ with the 
formula ${\cal E}=(E_{GB}-E_{PC})N/2A$, where $E_{GB}$ is the mean energy per particle of the system
in presence of grain boundaries, $E_{PC}$ is the same quantity computed for a perfect crystal in the
same simulation box, $A$ is the grain boundary surface and $N$ is the number of particles.
The results are showed in Tab.~\ref{tabella}: we notice that the surface energy of the interface 
increases with the density of the crystal and, at least at the lowest densities, ${\cal E}<2E_{LC}$,
being $E_{LC}$ the surface energy of the liquid-crystal interface\cite{BalLS, pederiva}. 
Our results are therefore in agreement with the experimental evidence of the stability of grain 
boundaries under the condition of phase coexistence between a crystal and a liquid\cite{Balibar} 
and confirm the result obtained from other PIMC calculations at finite temperature\cite{ProkGB}.

We also studied how the SGB system behaves in presence of vacancies. 
We have found that the activation of a vacancy in the crystal reduces the surface energy of the 
interfaces (see Tab.~\ref{tabella}). 
This result can be explained supposing that the vacancy are easily adsorbed into the grain boundary
contributing in relaxing the mechanical stress: the energy gain due to this adsorption process 
increases with the density. 
These simulation were carried on at fixed lattice constant; the activation of vacancies in the SGB 
system slightly reduces its average density and we found that, depending on the starting density, 
the activation of vacancies can destabilize the grain boundaries; such cases are indicated as GB 
unstable in Tab.~\ref{tabella}.

\begin{figure}[t]
 \label{condensato_gb}
 \centerline{\psfig{file=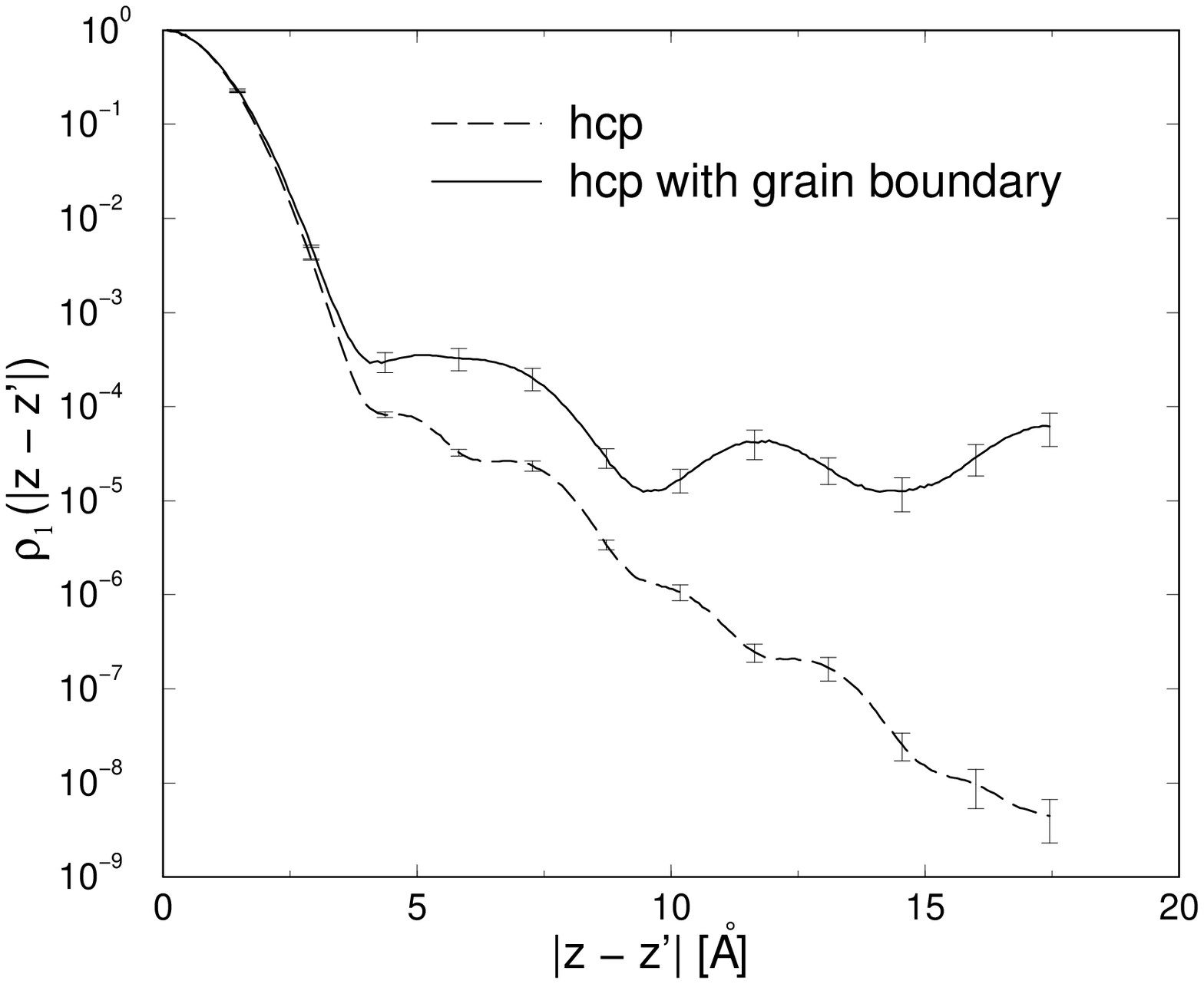,height=5.5cm}{(a)}
             \psfig{file=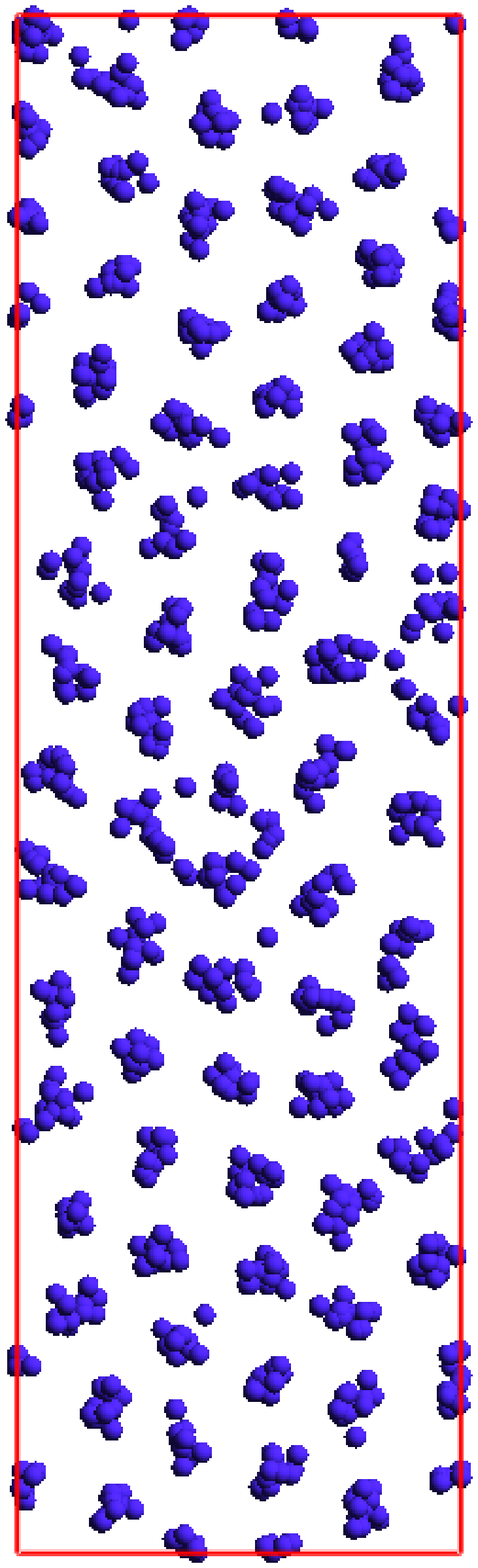,height=5.5cm}{(b)}}
 \caption{(a) One body density matrix computed at the density $\rho = 0.0313$ \AA$^{-3}$ along a
          direction orthogonal to the basal planes (z-axis) of a hcp commensurate lattice (dashed
          line) and along the same direction in the SGB system.
          (b) snapshot of a configuration of the polymers inside a basal plane of the SGB system;
          the calculation of $\rho_1$ is along the interfacial regions which are perpendicular to
          this basal planes.}
\end{figure}
Finally, in these systems, we computed the one body density matrix $\rho_{1}^{GB}(|z-z'|)$ at 
$\rho = 0.0313$ \AA$^{-3}$ along a direction parallel to the grain boundary strictly inside the 
interfacial region.
Our results are shown in Fig.~\ref{condensato_gb}; the $\rho_{1}^{GB}(|z-z'|)$, computed in the SGB
system, starts to oscillate around a non-vanishing value at a distance around 10 \AA.
The one-body density matrix computed in a commensurate hcp crystal along the same direction 
orthogonal to the basal planes is a exponentially decaying function of $|z-z'|$ in the range we have
studied.
This seems to suggest that, at zero temperature, also highly symmetrical grain boundaries are 
defects which can induce in the crystal off-diagonal long range order.
At the density $\rho = 0.0313$ \AA$^{-3}$, we found that the condensed fraction $n_{0}$ along these
interfaces is of the order of $3 \times 10^{-5}$. 
The tail of $\rho_{1}^{GB}(|z-z'|)$ shows oscillations related to the crystalline structure of the 
sample: these oscillations match the periodicity of the basal planes and the maxima represent the 
distance between two corresponding hcp basal planes.

\section{Conclusions}

The evidence is that even if it is possible that supersolid features observed in torsional 
oscillator experiments find their origin in some non equilibrium phenomena present in the solid 
$^4$He samples, the microscopic studies on this system, by means of Quantum Monte Carlo methods at 
zero and finite temperature, indicate how subtle is the barrier that separate this system from a 
macroscopic manifestation of the quantum nature which governs its behavior; it is sufficient to
introduce some defects to induce off-diagonal long range order in the system at least locally.
Vacancies seems to be the only kind of defects able to induce the coherence of the whole system.
The task for the future will be the explanation of the essential mechanism which underlie NCRI in
torsional oscillator experiment. 
Quantum Monte Carlo methods, specific to quantum solids with disorder, will play a leading role to 
attack this very interesting problem.


\end{document}